# Photoinduced Modulation of Rectification and Charge Transport in Diarylethene-Viologen Molecular Junctions

*Sofyan Tahri [1], Samuel Decorps [2], Stephane Lenfant [1], Denis Frath [2], and Imen Hnid [1]**

[1] Institute for Electronics Microelectronics and Nanotechnology (IEMN), CNRS, University of Lille, F-59652 Villeneuve d'Ascq, France.

[2] CNRS, ENS de Lyon, LCH, UMR 5182, 69342, Lyon cedex 07, France.

* Corresponding author: imen.hnid@iemn.fr

**KEYWORDS:** Molecular junctions; Diarylethene; Viologen; Photochromism; Rectification; Charge transport; Photocurrent

**ABSTRACT.** Large-area molecular junctions combining photochromic and redox-active functionalities provide an attractive platform for controlling charge transport by light. In this work, we study Au/DAEVIO/Ti/Au molecular junctions incorporated with electro-grafted layers of diarylethene-viologen with thicknesses of 12-16 nm. In the initial OFF state, the junctions exhibit low current densities and strong rectification, with a rectification ratio of 300 at ± 1V. Temperature-dependent measurements reveal a thermally assisted contribution at higher temperatures and a much weaker temperature dependence below 160-180K, consistent with multistep charge transport through molecular and/or interfacial states rather than direct coherent tunneling across the entire molecular layer. Under continuous in situ 385 nm irradiation, the electrical response is strongly and asymmetrically modified. The current density at +1V increases by a factor of 200, while the negative-bias current is only weakly modified, reducing the rectification ratio from 300 to 2. Illumination also produces a nonzero short-circuit current density of + 0.006mA $cm^{-2}$ and an open-circuit voltage of -0.45V. These results indicate that irradiation simultaneously modifies molecular charge transport and generates a measurable photoelectric response. This behavior is consistent with light-induced changes in electronic coupling associated with DAE photocyclization, combined with photoinduced charge generation and field-dependent charge separation involving the redox-active viologen units and asymmetric electrode interfaces.

**INTRODUCTION.** Molecular junctions (MJs), in which a molecular layer is sandwiched between two electrodes, provide a platform for investigating charge transport at the molecular scale and exploring electronic functionalities arising from molecular structure and properties.[1,2] In large-area solid-state junctions, charge transport depends not only on the electronic structure of the molecules but also on the thickness of the molecular layer and its coupling to the metallic electrodes.[3,4] For interelectrode distances below approximately 5 nm, electron transport is commonly described by direct off-resonant tunneling, characterized by an exponential decrease in current with increasing molecular thickness.[5,6] At larger distances, however, direct tunneling across the entire molecular layer becomes increasingly unlikely, and charge transport can involve multistep processes through molecular or interfacial states, including thermally assisted hopping and sequential tunneling.[5,7] Weakly temperature-dependent transport has also been reported in relatively thick molecular junctions, highlighting the diversity of transport regimes that can arise in these systems.[8,9]

The incorporation of redox-active molecular units offers an approach to modifying charge transport through electronically accessible molecular states. Viologens are particularly interesting in this context because of their reversible reduction processes and low-energy acceptor states, which can facilitate electron injection and transport.[10,11] Previous studies of viologen-based molecular junctions have shown that electronic coupling within the molecular layer and at the metal-molecule interfaces can strongly influence current density and rectification, even when the electrodes are intrinsically asymmetric.[11,12] These observations emphasize the importance of molecular connectivity in determining the electrical response of redox-active junctions. Introducing a photochromic unit into such a system offers the possibility of modifying this connectivity through light-induced changes in molecular structure.

Diarylethenes (DAEs) are photochromic molecules that undergo reversible transformations between open and closed forms upon irradiation at appropriate wavelengths.[13,14] This structural change modifies the π-conjugation of the molecular backbone and can consequently influence electronic coupling and charge transport.[15] Diarylethene-based molecular junctions have previously demonstrated light-induced conductance switching.[16,17] However, when a photochromic unit is combined with redox-active acceptor states in a relatively thick molecular layer, the electrical response may involve both changes in molecular connectivity and photoinduced charge generation and separation. Distinguishing these contributions is particularly relevant in metal/molecule/metal junctions, where the internal potential profile can influence charge injection, transport, and photocurrent collection.[18]

Here, we investigate solid-state large-area molecular junctions (MJs) incorporating a diarylethene-viologen molecule (DAEVIO) as active layers electrografted between Au and Ti/Au electrodes, with thicknesses of 12-16 nm, well beyond the distance range typically associated with direct coherent tunneling across the entire molecular layer. The initial electrical state of the junction is characterized by a low current density and strong rectification, with a rectification ratio RR(OFF) = 300 at ± 1 V. Temperature-dependent measurements reveal a thermally assisted contribution at high temperature and a much weaker temperature dependence below 160-180 K. Under continuous irradiation at 385 nm, the electrical response changes strongly and asymmetrically: the current at +1 V increases significantly, while it remains unchanged at -1V. Consequently, the rectification ratio decreases to $RR(ON_{h\nu})$ = 2 at ± 1 V and an ON/OFF ratio of R(ON/OFF) = 200 is obtained at +1V. More importantly, illumination produces a nonzero short-circuit current density, $J_{SC}$ = +0.006 mA $cm^{-2}$, together with an open-circuit voltage, $V_{OC}$ = -0.45 V.

These results show that irradiation modifies both the bias-dependent conductance and the photoelectric response of DAEVIO junctions, providing insight into the interplay between photochromism, redox-active molecular states, and asymmetric electrode interfaces.

DAEVIO combines a photochromic diarylethene unit (DAE), an aniline group (NH2-phenyl) and a phenyl-viologen acceptor (Ph-VIO) within the same molecular structure (**Figure 1a**). The synthesis process of this molecule will be reported separately. The molecular design was chosen to associate two complementary electronic functions. The DAE unit undergoes light-induced conversion between an open, weakly conjugated form and a more conjugated closed form (**Figure 1b**), whereas the phenyl-viologen unit introduces low-energy acceptor states into the molecular structure that promote efficient electron injection. The aniline groups were used for in situ diazonium generation and layer deposition[19,20] on gold strips as bottom electrodes, resulting in molecular layers with thicknesses of 12-16 nm (**Figure 1c**). Thin top electrodes consisting of 2nm of Ti and 10 nm of Au were then evaporated to complete the junction (**Figure 1d**). Details of the electrode fabrication process are provided in the Supporting Information. An LED source operating at 385 nm was used to irradiate the molecular junctions in situ (**Figure 1d**).

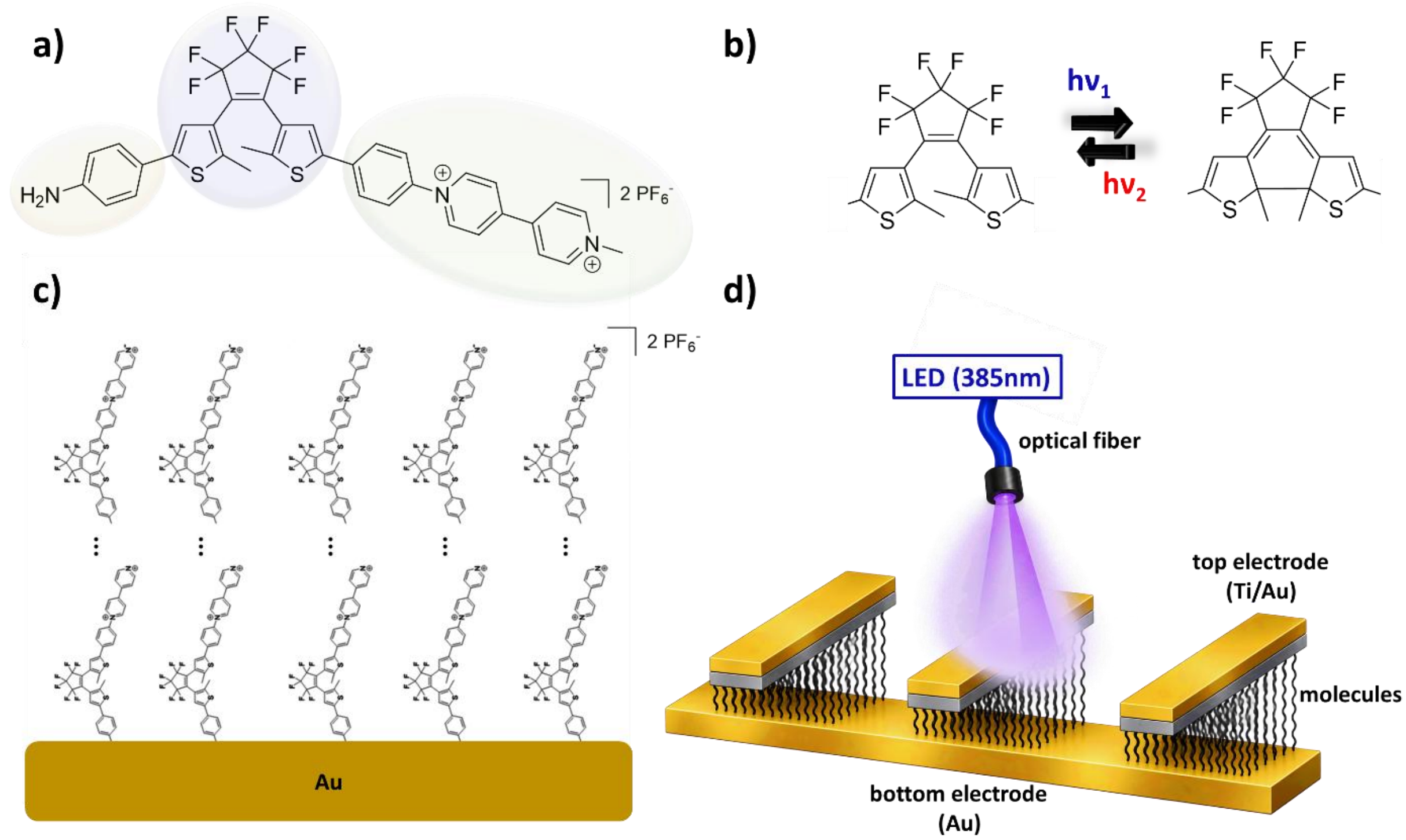

**Figure 1.** Illustration of the molecular junctions and experiments. (a) Molecular structure of NH2-phenyl-DAE-phenyl-VIO. (b) Photoisomerization of the diarylethene unit. (c) Idealized representation of the molecular layer obtained after electroreduction of the corresponding diazonium salt on the Au bottom electrode (50 nm). Note that the DAEVIO units within the electrografted layer are not as ordered as depicted in this idealized representation. (d) Scheme of experiment. The large-area molecular junction consists of the DAEVIO layer sandwiched between the Au bottom electrode and the Ti (2 nm)/Au (10 nm) top electrode. Light from the 385 LED is focused onto the molecular junction through the semitransparent top electrode (see **Section 4** of the Supporting Information for details).

**RESULTS AND DISCUSSION.** The photochromic properties of DAEVIO were first investigated in solution by UV-visible spectroscopy and cyclic voltammetry (**Figure S1a**). Irradiation at 385 nm produced the spectral changes expected for formation of the closed DAE form (**Figure S1a, red curve**). Subsequent irradiation at 455 nm induced the reverse spectral changes (**Figure S1a, blue dashed curve**), confirming the reversibility of the photochromic process and showing that the photochromic functionality is preserved in the DAEVIO molecule. As extensively documented, viologen cores are well-known redox-active units capable of reversibly exchanging two electrons.[21] The electrochemical behavior of DAEVIO was investigated by cyclic voltammetry before and after irradiation (**Figure S1b,c**). Two reduction processes characteristic of the viologen core are observed (**Figure S1b**), corresponding to its successive one-electron reductions. In contrast, irradiation mainly affects the oxidation region of the cyclic voltammogram (**Figure S1c**), consistent with a modification of the DAE electronic structure upon photocyclization. From the electrochemical potentials, the frontier molecular energy levels of DAEVIO were estimated (**Table S1**). The reduction-derived acceptor level, predominantly associated with the viologen unit, remains nearly unchanged upon irradiation at approximately -4.1 eV. In contrast, the oxidation-derived occupied level exhibits a pronounced shift from approximately -5.64 eV in the open form to -5.14 eV in the closed form. This 0.50 eV shift is consistent with the restoration of π-conjugation upon DAE ring closure and results in a substantial decrease in the estimated electrochemical gap. These results show that photoisomerization predominantly modifies the occupied electronic levels of DAEVIO while preserving the energetically accessible viologen-derived acceptor states.

To enable electrode surface functionalization, DAEVIO was converted *in situ* into its corresponding diazonium salt, followed by electrochemical reduction to graft the molecule onto the Au substrate. This well-established strategy for covalent surface modification[20,22] enables the formation of molecular films through the generation and coupling of aryl radicals. While this method provides robust anchoring, the electrochemical grafting conditions can be adjusted to control the film thickness, ranging from thin molecular layers to multilayered oligomeric films.[20,23] The diazonium derivative of DAEVIO was generated *in situ* by adding an excess of *tert*-butyl nitrite (tBuONO) to the precursor solution (**Figure S2a**). The electrochemical grafting was subsequently performed by cyclic voltammetry, where reduction of the diazonium groups generates aryl radicals that initiate covalent attachment to the Au surface and further intermolecular coupling within the growing film (**Figure S2b**). Typical grafting voltammograms are shown in **Figure S3**. During successive potential scans, the progressive decrease in the reduction current indicates increasing surface passivation as the DAEVIO layer grows on the electrode. By adjusting the grafting conditions, DAEVIO films with a thickness of 12 nm were obtained and subsequently used for the fabrication of Au/DAEVIO/Ti/Au molecular junctions. Such thicknesses are well beyond the distance range typically associated with direct coherent tunneling across the entire molecular layer.

To evaluate the molecular thickness, the molecular films were deposited on planar gold stripes using the same electrochemical conditions as described above and characterized by Atomic Force Microscopy (AFM) following a well-established method.[24] As shown in **Figure 2a** and **Figure S4** and detailed in **Table S2**, thicknesses around 12-16 nm were obtained. AFM was also used to evaluate the roughness of the grafted films, showing homogeneous molecular films with surface roughness values close to those of the bare gold electrode (**Figure S5**).

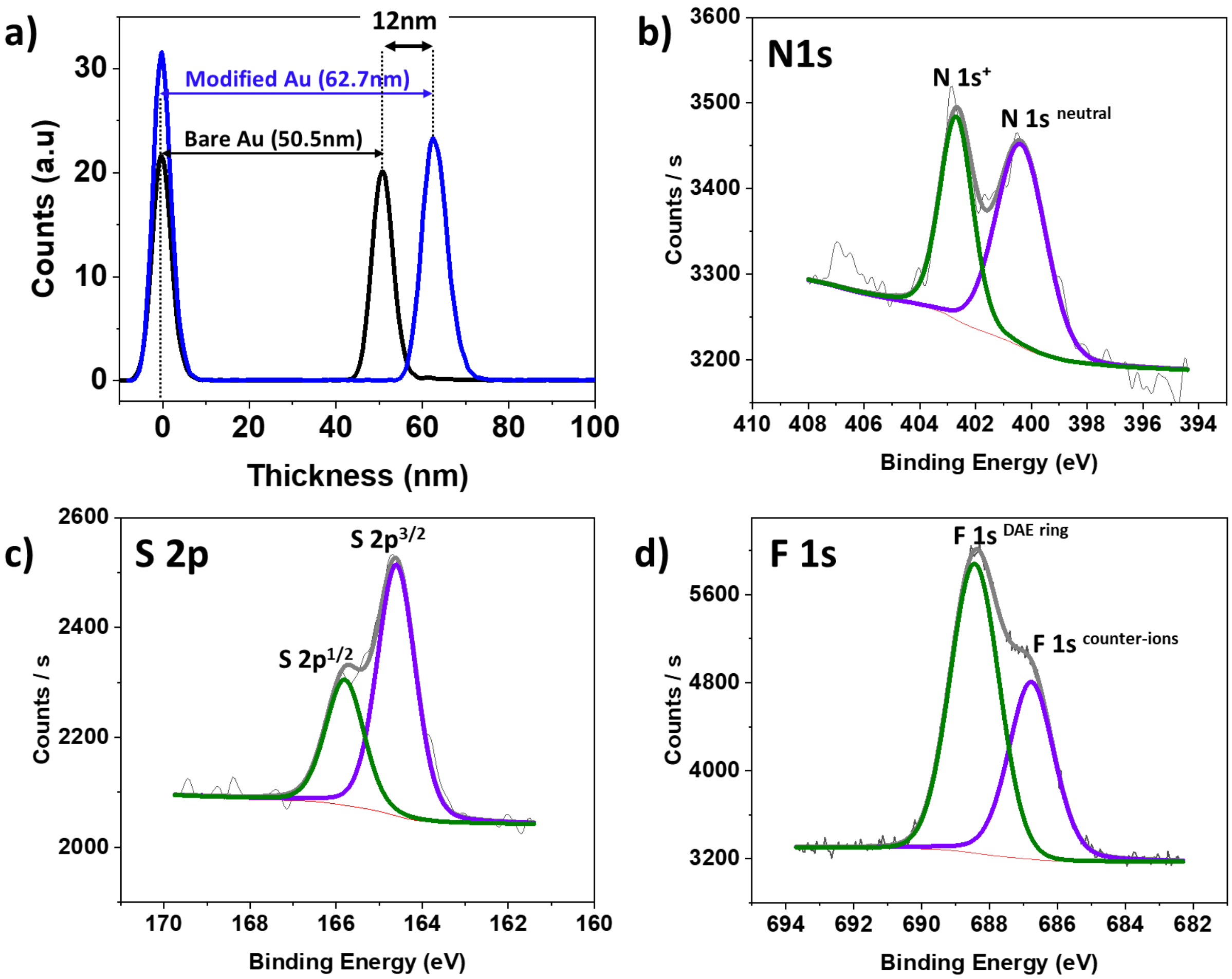

**Figure 2.** (a) Gaussian statistical analysis of the AFM height distributions, showing a DAEVIO film thickness of 12 nm. High-resolution XPS spectra of the DAEVIO film: (b) N 1s, (c) S 2p, and (d) F 1s.

To further characterize the grafted films, molecular films were deposited on continuous gold substrates using the same electrochemical conditions as described above and characterized by X-ray photoelectron spectroscopy (XPS). Before XPS analysis, the film thickness on these substrates was determined using the AFM scratching method, in which a small area of the molecular layer was mechanically removed with the AFM tip and the resulting step height was measured.[20] The measurements give a film thickness of 7 nm (**Figure S6**). This lower thickness is consistent with the larger grafted area compared with that of the gold stripe electrodes. The survey XPS spectrum confirmed the presence of the chemical constituents expected from the DAEVIO molecular structure and its associated $PF_6^-$ counterions (**Figure S7a**). The high-resolution Au 4f spectrum shows an attenuated Au $4f^{7/2}$ signal at approximately 84 eV compared with that of bare gold (**Figure S7b**), consistent with the formation of a molecular film on the surface and with a film thickness below approximately 10 nm, within the typical XPS information depth. High-resolution N 1s spectra (**Figure 2b**) reveal two components: a peak at 402.6 eV, attributed to positively charged nitrogen ($N^+$) of the viologen moiety, and a second at 400.3 eV, corresponding to neutral nitrogen. The experimental $N^+/N$ ratio of 0.5, together with an $N^+/N^{neutral}$ ratio close to 1, is consistent with the value expected for the singly reduced viologen state ($VIO^{\bullet+}$) (**Table 1**). The high-resolution C 1s spectrum (**Figure S8**) was deconvoluted into a main component at approximately 285.5 eV, assigned to C-C/C-H carbon atoms of the molecular backbone, and additional components at 286.5, 288.2, and 291.1 eV, attributed to C-N/C-S, $C=N^+$, and C-F environments, respectively. The experimental C/S ratio of 19.6 is in good agreement with the theoretical value of 19, while the N/S ratio of 1.2 is also close to the expected value of 1, supporting the expected composition of the grafted DAEVIO layer (**Table 1**). In addition, the S 2p spectrum exhibits the characteristic thiophene sulfur doublet at 164.6 and 165.8 eV (**Figure 2c**). The experimental $F^{DAE\text{-}ring}/S$ ratio of 3.1 agrees well with the theoretical value of 3, supporting the preservation of the perfluorocyclopentene-thiophene DAE unit after electrografting. Finally, the F 1s spectrum displays two contributions at 688.4 and 686.8 eV, assigned to the C-F groups of the DAE core and the PF6- counterions, respectively (**Figure 2d**). The experimental $F^{counter\text{-}ion}/P$ ratio of 5.8 is in excellent agreement with the theoretical value of 6, confirming the presence of PF6- counterions within the molecular film.

**Table 1.** Comparison between the theoretical and experimental atomic ratios determined from the corrected XPS peak areas.

| | **Theoretical ratio ($VIO^{2+}$)** | **Theoretical ratio ($VIO^{•+}$)** | **Theoretical ratio ($VIO^{0}$)** | **Experimental ratio** |
|---|---|---|---|---|
| **$N^{+}$/N** | 1 | **0.5** | 0 | **0.5** |
| **$N^{+}/N^{neutral}$** | - | **1** | 0 | **1.0** |
| **C/S** | 19 | **19** | 19 | **19.6** |
| **N/S** | 1 | **1** | 1 | **1.2** |
| **$F^{DAE\text{-}ring}$/S** | 3 | **3** | 3 | **3.1** |
| **$F^{counter\text{-}ions}$/P** | 6 | **6** | - | **5.8** |

Large-area MJs were subsequently fabricated by depositing thin Ti/Au top electrodes on the electrografted layers on the gold stripe bottom electrodes (**Figure 1d**). The junction architecture used for the main electrical investigation was Au(50nm)-DAEVIO(12nm)/Ti(2nm)/Au(10nm), with an active junction area of 20 x 20 $\mu m^2$. The thin Ti/Au top electrode was also sufficiently transparent at 385 nm to allow irradiation of the molecular layer, with a measured optical transmission of 34.3 % at this wavelength (**Table S4**).

The electrical response was investigated in two states. The initial state before irradiation is denoted OFF, whereas the response measured under continuous 385 nm irradiation is denoted $ON_{hv}$. Additional experiments involving prolonged 385 nm irradiation followed by measurements in the dark, leading to a persistent $ON_{dark}$ state, and subsequent irradiation at 455 nm, leading to an $OFF_{dark}$ state, will be presented in a separate study and are therefore beyond the scope of the present work.

**Figure 3** shows the current density-voltage characteristics of the Au(50nm)-DAEVIO (12nm)/Ti(2nm)/Au(10nm) molecular junctions before irradiation. In total, 36 junctions from 3 independently fabricated samples were electrically characterized. Among them, 31 exhibited stable electrical characteristics without short-circuit-like behavior, corresponding to a device yield of approximately 86%.

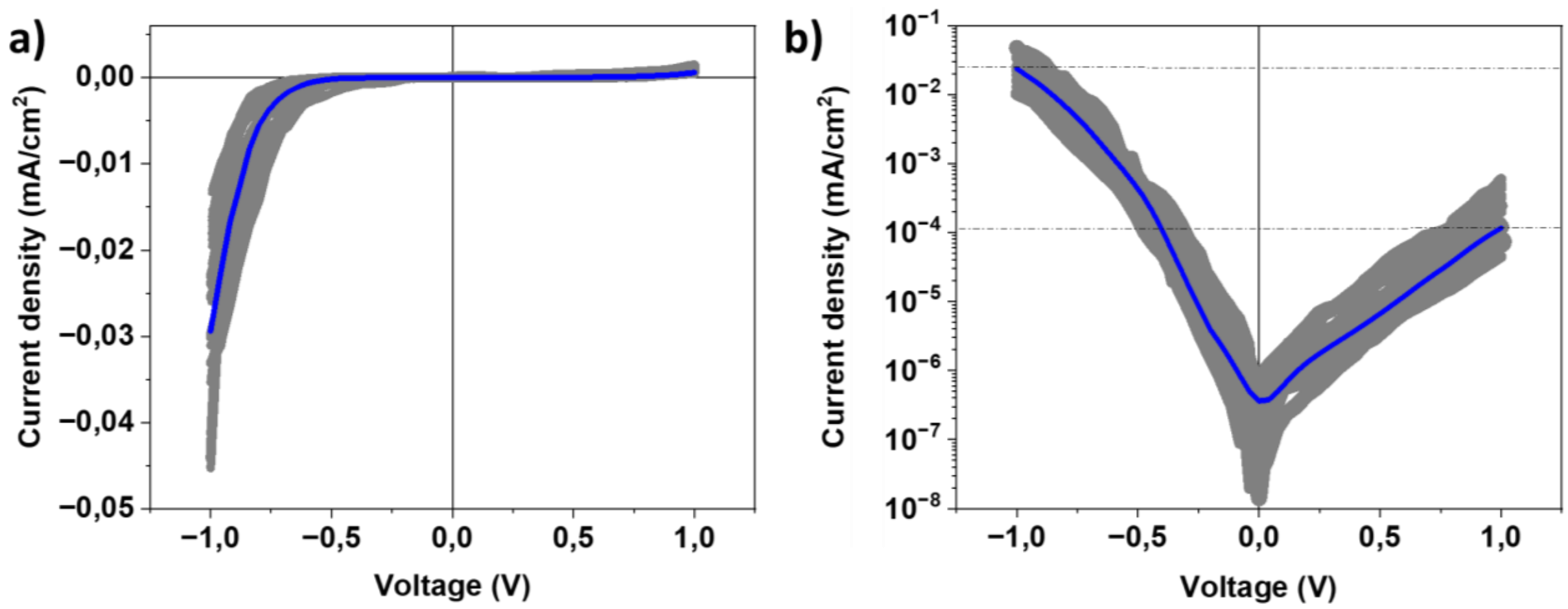

**Figure 3.** Electrical properties of the Au(50nm)-DAEVIO (12nm)/Ti(2nm)/Au(10nm) large-area molecular junctions at the initial state. (a) J-V and (b) log(|J|)-V characteristics. The grey curves correspond to the individual J-V measurements recorded on the 62 sweeps obtained from 31 electrical functional junctions (3 samples), while the blue curves represent the average J-V characteristic calculated from all measurements.

The functional junctions showed reproducible J-V characteristics, with a relatively limited dispersion between devices. At +1 V, the average current density was $10^{-4}$ mA cm$^{-2}$, whereas at -1 V its magnitude reached approximately 3*$10^{-2}$ mA cm$^{-2}$. The initial DAEVIO junction is therefore both weakly conducting and strongly rectifying and a rectification ratio of 300 is obtained. The low conductance of this state is consistent with the molecular structure of open

DAEVIO. Before irradiation, the DAE units are predominantly in their open, cross-conjugated form. Although the viologen unit provides low-energy acceptor states, electronic communication between successive molecular sites is limited by the open DAE segments.

The observed rectification indicates that charge transport is not equivalent for the two bias polarities. Several sources of asymmetry are present simultaneously in the junction: the bottom and top electrodes are different, the Au-DAEVIO and DAEVIO/Ti interfaces have different coupling conditions, and the DAEVIO molecular structure itself is asymmetric. The voltage drop across the junction may therefore also be distributed differently according to the polarity of the applied bias. Similar effects have been observed in viologen-based junctions, where modification of the electronic coupling at one electrode can transform an approximately symmetric junction into a strongly rectifying one.[12]

To obtain additional information on the transport regime in the initial state, the temperature dependence of the current was investigated. Temperature-dependent J-V measurements were performed on Au-DAEVIO/Ti/Au junctions containing a 16 nm DAEVIO film (grafting conditions and thickness measurement are given in **Table S2** and **Figure S4f**, respectively). These junctions were investigated during cooling from 300 to 77 K and subsequent warming to 300 K (**Figure S11**). The magnitude of the current decreased progressively upon cooling. The temperature dependence was more pronounced above approximately 160-180 K and became considerably weaker at lower temperature. The J-V characteristics measured during warming closely recovered those obtained during cooling. The rectifying shape was also maintained over the investigated temperature range. No major irreversible modification of the electrical response was therefore detected during the thermal cycle. The temperature dependence was analyzed at +2 and -2 V using $|J| = J_0 \exp\left(-E_a/(k_B T)\right)$ (**Figure S12**). At both polarities, the nominal average electric-field magnitude is approximately 0.125 V $nm^{-1}$. The Arrhenius plots show two distinct slopes. Between 180 and 300 K, apparent activation energies of 52-54 meV were obtained at +2 V and 91-102 meV at -2 V. Between 77 and 160 K, the corresponding values decreased to 3-4 and 7-8 meV, respectively (**Figure S12**).

The high-temperature activation energies indicate a thermally assisted contribution to charge transport. Considering the 16 nm molecular thickness, this behavior is compatible with multistep transport through molecular or intermolecular states, potentially involving thermally assisted hopping and sequential electronic-transfer processes.[5,8]

Below approximately 160-180 K, the much smaller apparent activation energies indicate that transport becomes only weakly temperature dependent. Similar transitions from thermally activated to weakly temperature-dependent transport have been observed in relatively thick molecular junctions.[25] A qualitatively similar behavior has also been reported for viologen-based Au-VIO/Ti/Au junctions, where the apparent activation energy decreased strongly at low temperature.[9]

Interestingly, the activation energy also depends on bias polarity. In the electrical configuration used for the temperature-dependent measurements, the more conductive branch exhibits the lower Ea, whereas the less conductive branch exhibits the higher Ea (**Figure S12**). Because the magnitude of the nominal applied field is the same, this difference points to polarity-dependent effective transport conditions. The asymmetric Au-DAEVIO and DAEVIO/Ti interfaces can produce different injection conditions and potential distributions for the two bias directions, consistent with the rectification observed. Overall, these results support a multistep transport picture involving molecular and/or interfacial states.

The molecular junctions were subsequently investigated under continuous irradiation at 385 nm. **Figure 4** shows the current density-voltage characteristics of the Au-DAEVIO/Ti/Au molecular junctions under continuous illumination. Compared with the OFF state, a pronounced modification of the electrical response is observed. At +1 V, the current density increases from $10^{-4}$ mA cm$^{-2}$ in the OFF state to $2 \times 10^{-2}$ mA cm$^{-2}$ under illumination, corresponding to an ON/OF ratio of R(ON/OFF) = 200. At -1 V, the current density changes much less, from -0.03 to -0.04 mA cm$^{-2}$. Consequently, illumination strongly suppresses the initial rectification: RR(OFF) = 300 and RR($ON_{h\nu}$) = 2. These results show a strong polarity dependence under the irradiation. Here, the initial low-conductance positive-bias branch is preferentially enhanced, while the negative-bias branch is only weakly modified.

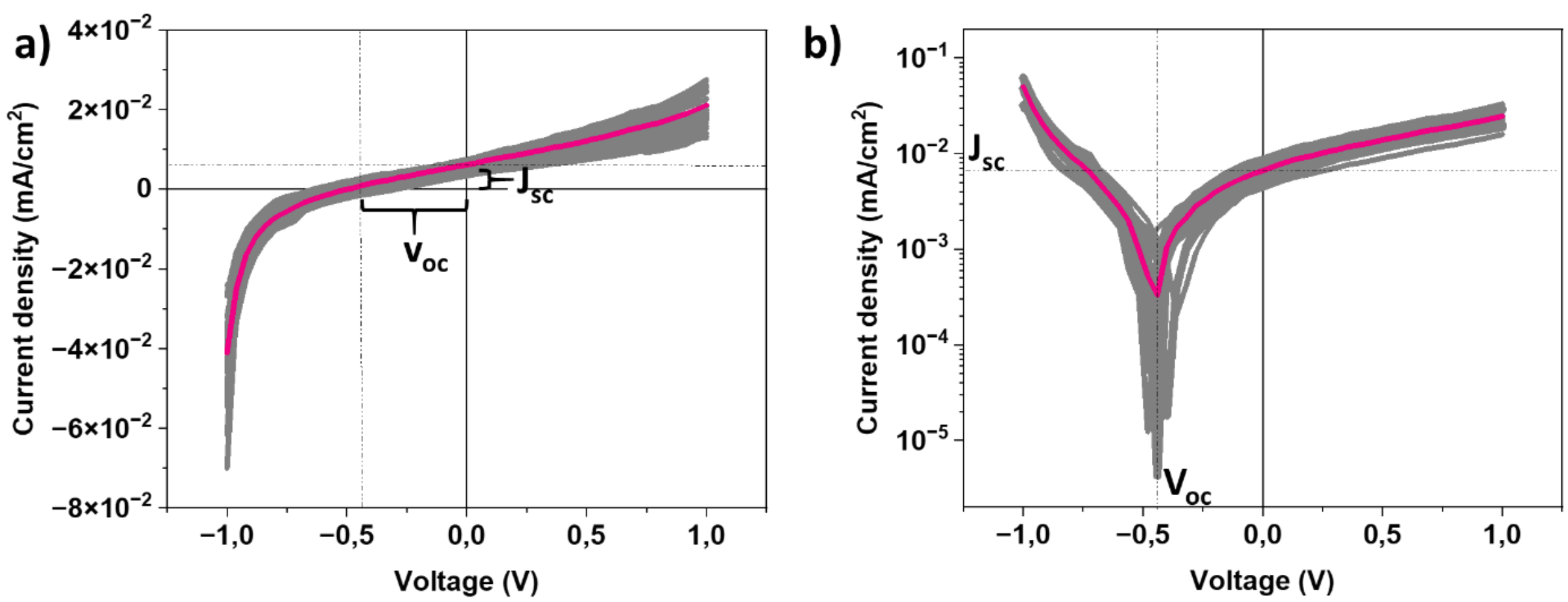

**Figure 4.** Electrical characteristics of the Au(50nm)-DAEVIO (12nm)/Ti(2nm)/Au(10nm) junctions under in-situ 385 nm irradiation. (a) J-V and (b) log(|J|)-V curves. The pink curves are the J-V average calculated from 62 measurements performed on three samples (grey curves). The low-bias region highlights the short-circuit current density (JSC, at V = 0 V) and the open-circuit voltage (VOC, at J = 0).

A second important characteristic appears around zero bias. Under continuous illumination, the J-V curve no longer passes through the origin. A short-circuit current density $J_{SC}$ = +0.006 $mAcm^{-2}$ is measured at zero applied voltage, together with an open-circuit voltage $V_{OC}$ = -0.45V. Both characteristics were reproducibly observed in the electrically functional junctions under illumination. All 31 electrically functional junctions exhibited an *in situ* photoinduced ON response under irradiation, corresponding to a 100% photoresponse yield among functional devices. The nonzero $J_{SC}$ shows that illumination produces a net electrical current without an externally applied bias. Conversely, $V_{OC}$ corresponds to the external voltage required to compensate this photoinduced current. The opposite signs of $J_{SC}$ and $V_{OC}$ are therefore consistent with the measured J-V response. Photovoltaic-like responses have previously been reported in photoactive large-area molecular junctions, where molecular orbital energies, asymmetric interfaces, and the internal potential profile influence photocarrier generation, separation, and collection.[1,18] A similar situation can be considered here. The different Au-DAEVIO and DAEVIO/Ti interfaces provide an asymmetric environment for charge injection and extraction, while excitation of the DAEVIO molecular layer creates a nonequilibrium electronic population.

The simultaneous observation of conductance modulation and nonzero $J_{SC}$ and $V_{OC}$ shows that the $ON_{h\nu}$ response cannot be described solely as the conductance change associated with DAE photocyclization. Under continuous irradiation, at least two effects can contribute: a modification of the molecular transport network associated with the conversion of open DAE units into the closed form, and an instantaneous photoelectric contribution associated with light absorption, charge separation, transport, and extraction.

The evolution of the rectification ratio provides an additional view of the effect of irradiation: RR(OFF) = 300, RR($ON_{h\nu}$) = 2. The electrodes remain physically unchanged during this sequence.

The disappearance of rectification therefore cannot result from removal of the structural asymmetry between Au and Ti/Au. Instead, it indicates that the electronic state of the molecular layer strongly controls how this interfacial asymmetry is expressed in the measured current. This behavior can be compared with previous viologen MJs.[11] Strongly coupled viologen junctions can exhibit approximately symmetric J-V characteristics, whereas decreasing the electronic coupling at one interface produces lower current and pronounced rectification.[12]

In the present DAEVIO junctions, photocyclization modifies both the internal electronic coupling of the molecular layer and its energetic relationship with the contacts. The preferential increase of the initially low-conductance positive-bias branch under continuous irradiation is consistent with such a modification. This does not imply that the Au-DAEVIO and DAEVIO/Ti interfaces become electronically equivalent. Rather, the complete junction evolves from an interface-sensitive, poorly connected transport network toward a more efficiently connected molecular state in which the initial electrical asymmetry is much less apparent.

**Figure 5** presents a plausible energy diagram for DAEVIO MJs based on the electrochemical energy levels and the electrical, temperature-dependent, and photoresponse measurements. At equilibrium and in the absence of light irradiation (**Figure 5a**), the difference in the work functions of the two metallic electrodes (Au, 5.1 eV; Ti, 4.3 eV) is expected to induce an internal potential gradient across the DAEVIO layer. In the OFF state, the open DAE units limit the electronic coupling between successive viologen-derived states, resulting in a series of localized LUMO states through which charge transport can occur by multiple hopping and/or tunnelling steps. At the Au-DAEVIO interface, the energy difference between the Au Fermi level and the first accessible DAEVIO LUMO introduces an injection barrier, $\Phi_1$, whereas at the DAEVIO/Ti interface, the terminal viologen-derived state lies close to the Ti Fermi level, resulting in a much smaller injection barrier. Under positive applied bias (**Figure 5c**), the resulting energy-level configuration remains unfavorable for electron injection and transport, leading to a relatively low current. In contrast, under negative bias (**Figure 5b**), the molecular energy profile becomes more favorable for charge injection and subsequent transport through the localized viologen-derived states, resulting in the higher current experimentally observed at negative bias. This asymmetric energy-level configuration therefore provides a plausible explanation for the strong rectification observed in the OFF state.

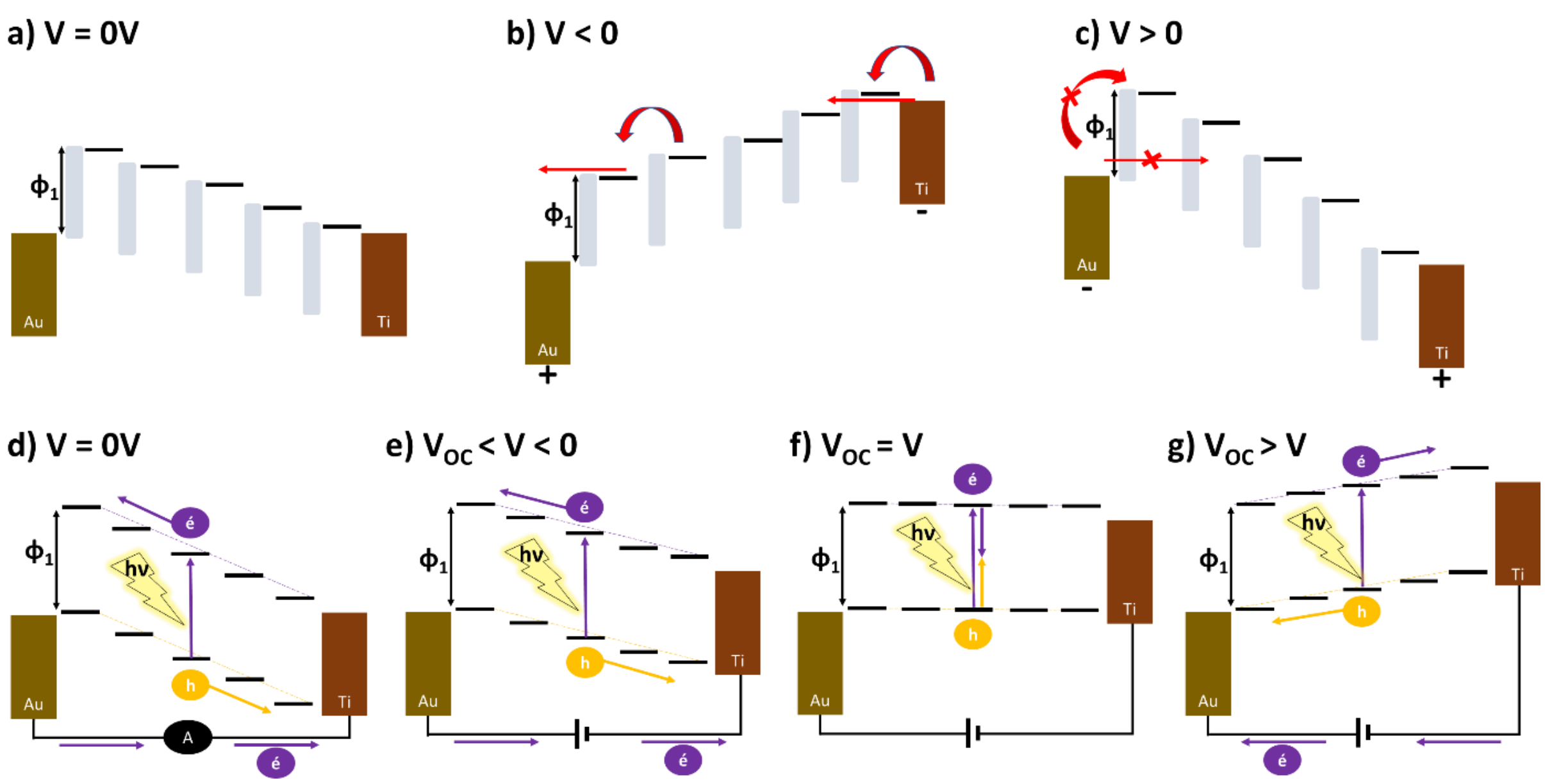

**Figure 5**. Proposed energy-level diagrams for charge transport and photoresponse in Au/DAEVIO/Ti/Au molecular junctions. (a-c) OFF-state configurations at (a) zero bias, (b) negative bias, and (c) positive bias. $\Phi_1$ represents the injection barrier at the Au/DAEVIO interface. Curved and straight arrows indicate multistep hopping and tunneling processes, respectively. (d-g) proposed photoresponse under irradiation at (d) V=0V, (e) $V_{OC}$<V<0, (f) V=$V_{OC}$, (g) V<$V_{OC}$.

Under irradiation (**Figure 5d**), photoexcitation of the DAEVIO layer generates excited charge carriers that can be separated and collected under the internal electric field, resulting in a finite photocurrent at zero applied bias. The magnitude and direction of this field are expected to depend not only on the asymmetric Au and Ti contacts but also on the electronic and ionic response of the DAEVIO layer. In particular, charge redistribution involving the redox-active viologen units, together with possible displacement of the associated counter-ions, may partially screen the internal electric field. Such an effect has previously been proposed for viologen-containing molecular junctions and could contribute to the photoresponse observed here.[26,27] When a bias is applied (**Figure 5e-g**), the external electric field modifies the internal potential gradient and consequently the separation and collection of the photogenerated charges, accounting for the bias-dependent photocurrent and its change in direction around $V_{OC}$. At $V = V_{OC}$ (**Figure 5f**), the applied external potential compensates the driving force responsible for the net photocurrent, resulting in zero net photocurrent. For applied voltages more negative than $V_{OC}$ (**Figure 5g**), the external field becomes dominant, resulting in a reversal of the photocurrent direction. This behavior is consistent with the experimentally observed non-zero $J_{SC}$ and $V_{OC}$ and indicates that the photoresponse of the DAEVIO junctions involves photoinduced charge generation together with field-dependent charge separation and transport.

**CONCLUSION:** In this study, we investigated charge transport and photoinduced electrical responses in large-area Au-DAEVIO/Ti/Au molecular junctions incorporating electrografted diarylethene–viologen layers. In the initial OFF state, junctions with a 12 nm-thick molecular layer exhibit low current densities and strong rectification, with a rectification ratio of 300 at ±1 V. Temperature-dependent measurements on 16 nm-thick junctions reveal a thermally assisted contribution at higher temperatures, with apparent activation energies of 52-54 meV at +2 V and 91-102 meV at -2 V, followed by a much weaker temperature dependence below approximately 160-180 K. These observations are consistent with multistep charge transport involving molecular and/or interfacial states rather than direct coherent tunneling across the entire molecular layer.

Under continuous 385 nm irradiation, the electrical response changes markedly. The current density at +1 V increases by a factor of 200, while the negative-bias current changes only slightly, reducing the rectification ratio from 300 to 2. In addition, a nonzero short-circuit current density of +0.006 mA cm-2 and an open-circuit voltage of -0.45 V are observed, demonstrating a photoelectric contribution beyond a simple change in molecular conductance. The combined results are consistent with a picture in which photocyclization of the DAE units modifies electronic coupling within the molecular layer, while photoinduced charge generation and field-dependent charge separation contribute to the response under illumination. The redox-active viologen units and asymmetric Au-DAEVIO and DAEVIO/Ti interfaces are expected to play important roles in these processes. By combining photochromic and redox-active functionalities within the same electrografted molecular layer, this work demonstrates that light can strongly modulate rectification while simultaneously generating a measurable photoelectric response in relatively thick molecular junctions. These findings highlight the potential of multifunctional molecular architectures for developing optically responsive molecular electronic devices.

ASSOCIATED CONTENT

**Supporting Information**.

The Supporting Information is available free of charge on the ACS Publications website. It includes materials and methods, UV-Vis, electrochemical deposition, AFM, XPS, light irradiation and IV curve analysis. (PDF)

AUTHOR INFORMATION

**Corresponding Author**

Imen Hnid - *Institute for Electronics Microelectronics and Nanotechnology (IEMN), CNRS, University of Lille, Av. Poincaré, Villeneuve d'Ascq, France*; ORCID: 0000-0002-4468-5334; Email: imen.hnid@iemn.fr

**Authors**

Sofyan Tahri - *Institute for Electronics Microelectronics and Nanotechnology (IEMN), CNRS, University of Lille, F-59652 Villeneuve d'Ascq, France*; ORCID: 0009-0007-[23]43-4425

Denis Frath - *CNRS, ENS de Lyon, LCH, UMR 5182, 69342 Lyon Cedex 07, France*; ORCID: 0000-0001-6888-9377

Stephane Lenfant - *Institute for Electronics Microelectronics and Nanotechnology (IEMN), CNRS, University of Lille, F-59652 Villeneuve d'Ascq, France*; ORCID: 0000-0002-6857-8752

**Author Contributions**

The manuscript was written through contributions of all authors. All authors have given approval to the final version of the manuscript. S.T. fabricated and characterized the MJs. D.F. synthesized the molecule. S.T and I.H. analyzed the I-V datasets. S.L. and I.H. supervised the project.

**Funding Sources**

University of Lille.

**Notes**

The authors declare no competing financial interest.

**ACKNOWLEDGMENT**

We acknowledge funding from the University of Lille and the technical support from the staff of IEMN: The Micro Nano Fabrication Centre (CMNF) and the scanning-probe microscopy characterization facilities (PCMP-PCP) are partly supported by the French Network Renatech. We thank X. Wallart (IEMN) for the XPS measurements, L. Thomas (IEMN) for the help of the AFM measurements.

**ABBREVIATIONS**

AFM, atomic force microscopy; DAE, diarylethene; DAEVIO, diarylethene-viologen; JSC, short-circuit current density; LUMO, lowest unoccupied molecular orbital; MJs, molecular junctions; RR, rectification ratio; VOC, open-circuit voltage; VIO, viologen; XPS, X-ray photoelectron spectroscopy.

# Photoinduced Modulation of Rectification and Charge Transport in Diarylethene-Viologen Molecular Junction

*Sofyan Tahri [1] Samuel Decorps [2] Stephane Lenfant [1] Denis Frath [2] and Imen Hnid [1]**

[1] Institute for Electronics Microelectronics and Nanotechnology (IEMN), CNRS, University of Lille, Av. Poincaré, Villeneuve d'Ascq, France.

[2] CNRS, ENS de Lyon, LCH, UMR 5182, 69342, Lyon cedex 07, France.

* Corresponding author imen.hnid@iemn.fr

## Supporting Information

# I. Materials and methods

**Solvents and Reagents:** Tetrabutylammonium hexafluorophosphate ($TBAPF_6$), used as the supporting electrolyte, and *tert*-butyl nitrite (tBuONO), used for the *in situ* formation of the diazonium species, were purchased from Sigma-Aldrich and used as received. All electrochemical solutions were freshly prepared in acetonitrile.

**UV-visible:** Absorption spectra were recorded in acetonitrile solution at room temperature. Absorption measurements over the spectral range from 280 nm to 800 nm are carried out in quartz cells having an optical path length of 1 cm.

**Electrochemistry:** Electrochemical experiments were performed using a PalmSens potentiostat in a three-electrode electrochemical cell. Measurements were carried out at 293 K in acetonitrile containing tetrabutylammonium hexafluorophosphate ($TBAPF_6$, 0.1 M) as the supporting electrolyte. Millimolar solutions of DAEVIO were used for the electrochemical characterization. Glassy carbon or Au electrodes were used as working electrodes for the solution-phase measurements, whereas 50 nm-thick Au stripes deposited on $Si/SiO_2$ substrates were used for the electrografting experiments. A platinum wire was used as the counter electrode and an Ag/AgCl electrode as the reference electrode. All electrochemical measurements were performed under an inert atmosphere after bubbling argon through the solutions prior to the experiments.

**Electrochemical Deposition of the DAEVIO Layers:** The diazonium derivative of DAEVIO was generated *in situ* using *tert*-butyl nitrite (tBuONO). Briefly, an acetonitrile solution containing DAEVIO and $TBAPF_6$ (0.1 M) as the supporting electrolyte was prepared and degassed by argon bubbling prior to the experiment. An excess of tBuONO was then added to the solution to convert the terminal amine groups into the corresponding diazonium species. After an incubation period, electrochemical grafting was performed using a PalmSens potentiostat in a three-electrode configuration, with the Au stripes as the working electrode, a platinum wire as the counter electrode, and an Ag/AgCl electrode as the reference electrode. The diazonium species were electrochemically reduced by successive potential sweeps, generating aryl radicals that resulted in covalent attachment to the Au surface and the progressive formation of oligomeric DAEVIO films. The grafting conditions were adjusted to obtain molecular layers with thicknesses of 12-16 ±1 nm. After electrografting, the modified Au substrates were removed from the grafting solution, thoroughly rinsed with acetonitrile, and sonicated in acetonitrile to remove weakly adsorbed species, before being dried and used for further characterization and molecular-junction fabrication.

**XPS Measurements:** X-ray photoelectron spectroscopy (XPS) was used to characterize the chemical composition of DAEVIO films electrografted onto Au substrates. High-resolution XPS spectra were acquired using a monochromatic Al Kα X-ray source (hν = 1486.6 eV), with a detection angle of 45° relative to the sample surface, an analyzer entrance slit width of 400 μm, and a pass energy of 12 eV. Under these conditions, an overall energy resolution of 0.55 eV was obtained, as determined from the full width at half-maximum (FWHM) of the Ag $3d^{5/2}$ line. A Shirley background was subtracted, and the photoemission peaks were fitted using Voigt functions and a least-squares minimization procedure.[1,2] Binding energies were referenced to the C 1s signal at 285.5 eV.

**AFM Measurements:** Atomic force microscopy (AFM) was used to characterize the surface morphology and determine the thickness of the electrografted DAEVIO films. All AFM measurements were performed in tapping mode to avoid damage to the molecular layer during scanning. Tapping mode AFM tips of Silicon (NCST, C=7.4N, fo=160kHz) from NANOWORLD were used. Film thickness was mainly determined using the lithographically defined gold-step method.[3] DAEVIO was electrografted onto 20 μm-wide Au stripes with a step height of 50 nm, deposited on $Si/SiO_2$ substrates. The molecular-layer thickness was determined from the difference between the statistical step heights measured for the bare and DAEVIO-functionalized Au regions. Height distributions were analyzed by Gaussian fitting using Gwyddion software. For continuous Au substrates, the film thickness was determined using the AFM scratching method, in which a small area of the molecular layer was mechanically removed with the AFM tip and the resulting step height was measured.

**Device Fabrication:** Molecular junctions were fabricated using established methods.[4] First, photolithography and metal evaporation were used to fabricate 50 nm-thick Au bottom electrodes with a width of 20 μm and a length of 1 cm on $Si/SiO_2$ substrates. DAEVIO molecular layers were then deposited onto the Au electrodes by electrochemical reduction of the corresponding *in situ*-generated diazonium salt. The electrografting conditions were adjusted to obtain oligomeric molecular layers with thicknesses around 12 ± 1 nm. A second photolithography/metal deposition step was subsequently performed to define the Ti/Au top electrodes and complete the crossed Au/DAEVIO/Ti(Au) molecular junctions, with an active junction area of 20 x 20 $\mu m^2$ (400 $\mu m^2$). The fabrication process showed good reproducibility: among 36 junctions fabricated across three samples, 31 exhibited functional electrical characteristics, corresponding to a device yield of approximately 86%.

**In Situ Irradiation Setup:** Light irradiation was provided by fiber-coupled LEDs (Thorlabs) operating at 385 (M385FP1) and 455 nm (M455FP1), coupled to a multimode optical fiber (BF20HSMA01, Thorlabs). The 385 nm LED was used to induce photocyclization of the DAE units toward the closed form, whereas irradiation at 455 nm was used to promote the reverse cycloreversion toward the open form. The output of the optical fiber was positioned close to the molecular junction, allowing direct illumination through the semitransparent Ti/Au top electrode (2 nm Ti and 10 nm Au). In this geometrical configuration, the thin metallic top electrode allowed partial transmission of 34.3 % (at 385nm) and 36.7% (at 455nm) of the incident light to the molecular layer (see **Table S4**). The transmitted optical power was measured using a Thorlabs PM204 power meter equipped with a silicon S120VC photodetector positioned at the sample location. All IV curves were first recorded in the initial state, i.e., in the dark. The junctions were then irradiated *in situ*, and the IV measurements were repeated under continuous illumination. Importantly, all 31 electrically functional junctions exhibited an *in situ* photoinduced ON response under irradiation, corresponding to a 100% photoresponse yield among functional devices.

**Electrical measurements:** Electrical measurements were performed using an Agilent 4156C semiconductor parameter analyzer, enabling *in situ* current-voltage measurements under controlled illumination.

## II. Characterization in solution

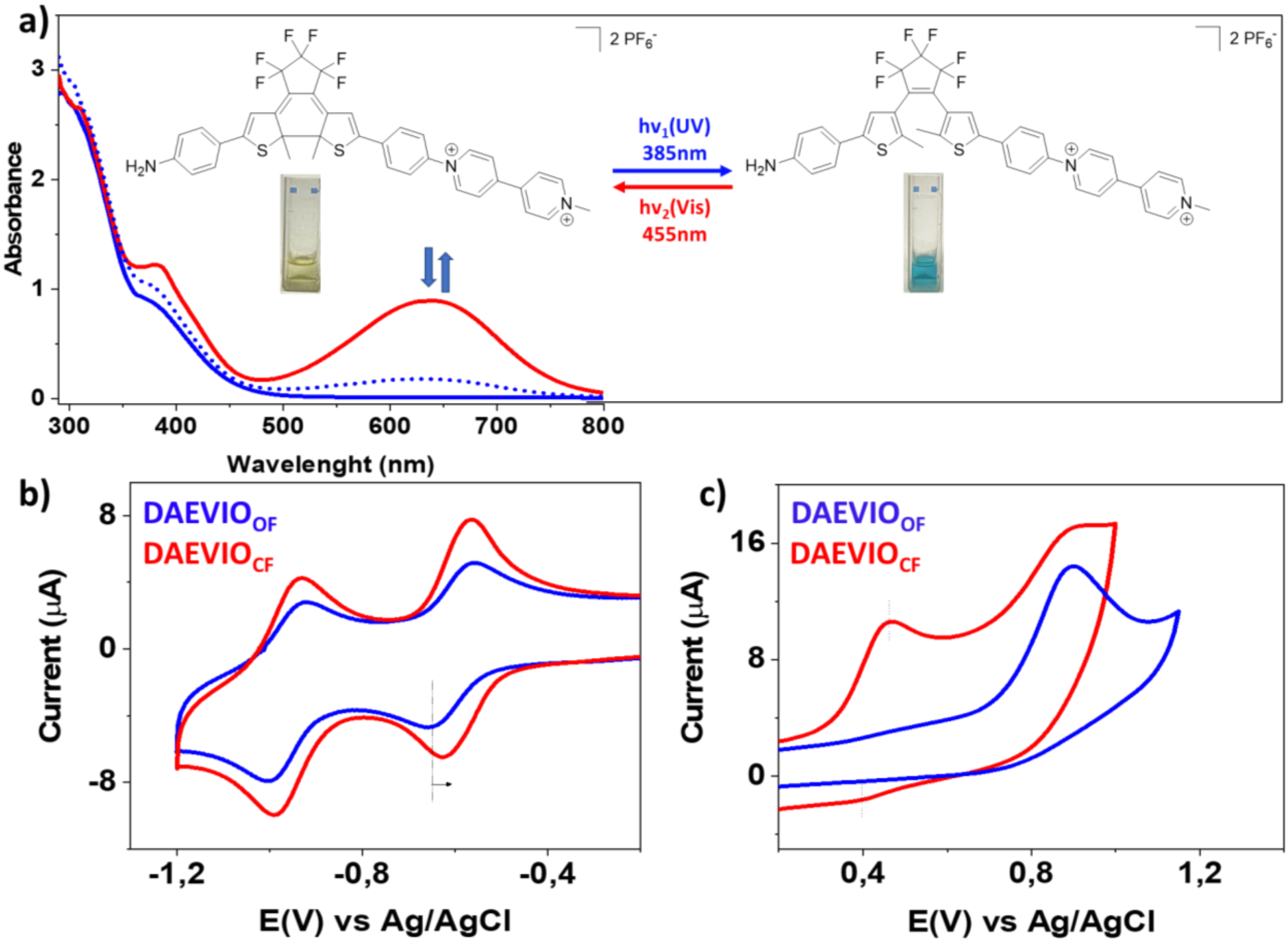

**Figure S1.** (a) UV-visible absorption spectra of DAEVIO (C = 0.5 mM) in acetonitrile: before irradiation (solid blue line) and after 30min of UV irradiation at 385 nm (solid red line). The dashed lines correspond to the spectra recorded after a few minutes of visible-light

irradiation (455 nm), demonstrating the partial reversibility of the photochromic switching. (b) and (c) Electrochemical behavior of DAEVIO (open and closed forms) in acetonitrile. Cyclic voltammograms of DAEVIO with the DAE unit in the open (blue) and closed (red) forms recorded in the (a) negative and (b) positive potential ranges. Supporting electrolyte: 0.1 mol.L$^{-1}$TBAPF$_6$, scan rate: 0.1 V.s$^{-1}$.

| | HOMO (eV) | LUMO (eV) | HOMO-LUMO gap (eV) |
|---|---|---|---|
| **DAEVIO$_{OF}$** | **- 5.64** | **- 4.05** | **1.6** |
| **DAEVIO$_{CF}$** | **- 5.14** | **- 4.09** | **1.0** |
| **DAE$_{OF}$** [1,2] | **- 5.49** | **- 2.98** | **2.5** |
| **DAE$_{CF}$** [1,2] | **- 5.15** | **- 3.48** | **1.7** |

**Table S1.** Experimentally determined HOMO and LUMO energy levels, and HOMO-LUMO energy gap, of DAEVIO obtained by electrochemistry, together with literature values for DAE[5,6].

# III. Electrografting

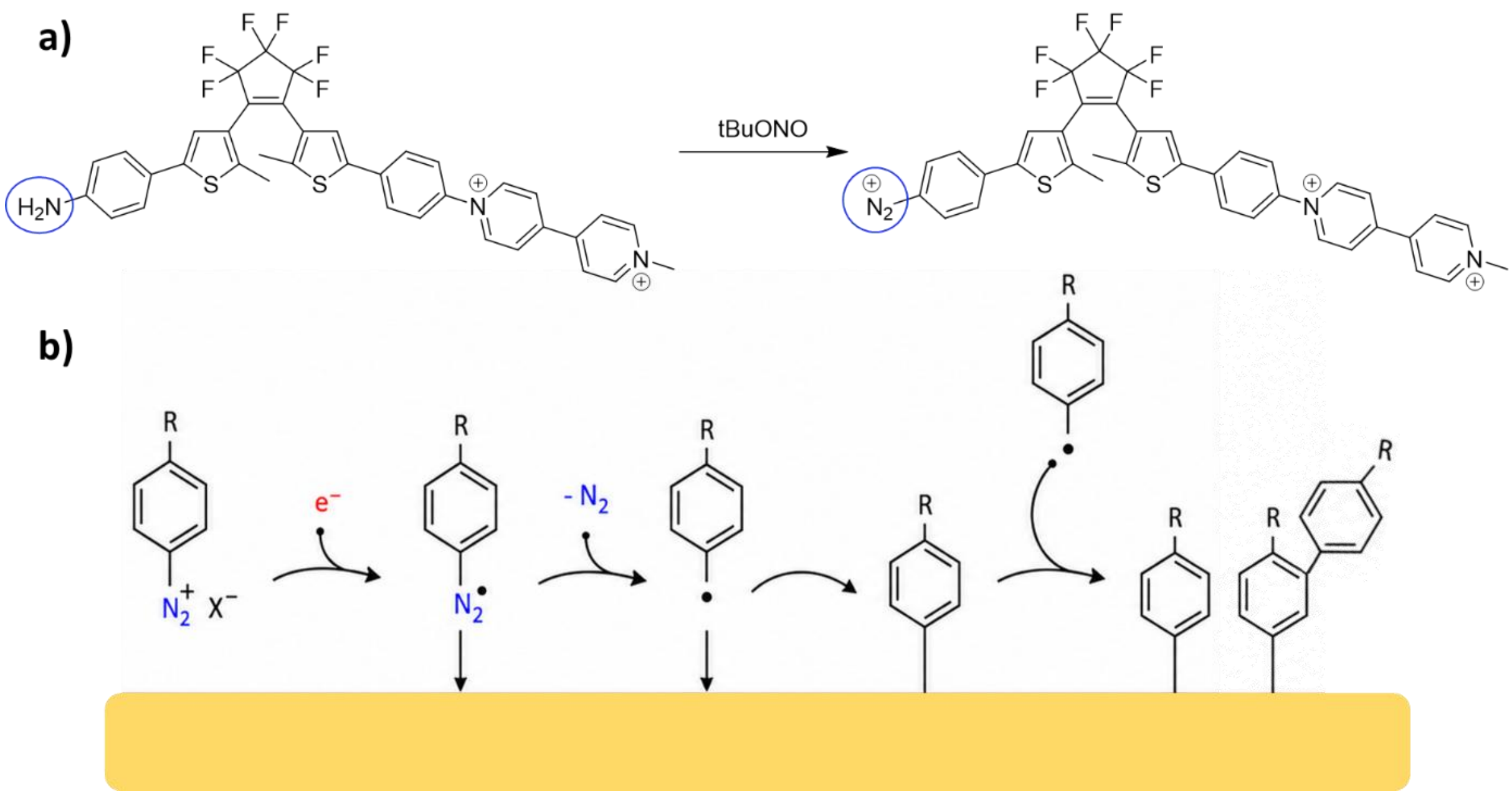

**Figure S2.** Schematic illustration of (a) diazonium salt electroreduction of the DAEVIO molecule, (b) electroreduction and grafting of diazonium species on a surface, adapted from[7].

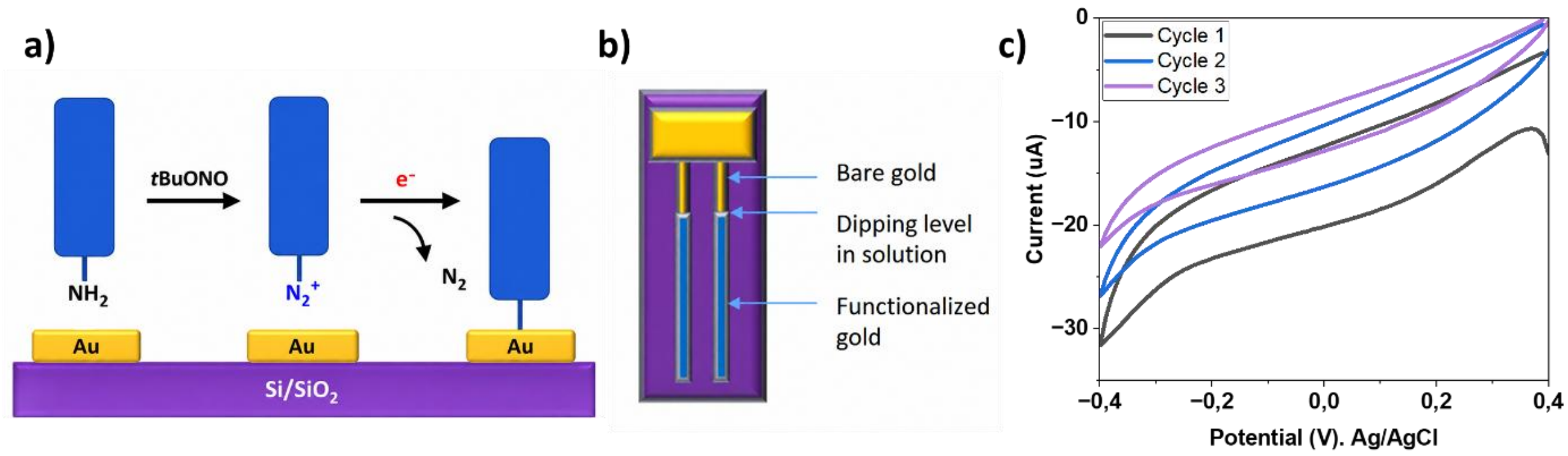

**Figure S3.** (a) Schematic illustration of electro-reductive grafting of diazonium salts onto a gold stripe. (b) Representation of a gold stripe used as a working electrode. (c) Cyclic voltammograms showing the reductive grafting process of diazoniums salt (5 mM) by sweeping potentials in three cycles between +0.4 V.Ag/AgCl and -0.4 V.Ag/AgCl on a continuous gold substrate.

# IV. Characterization in the solid state

## IV.1. AFM

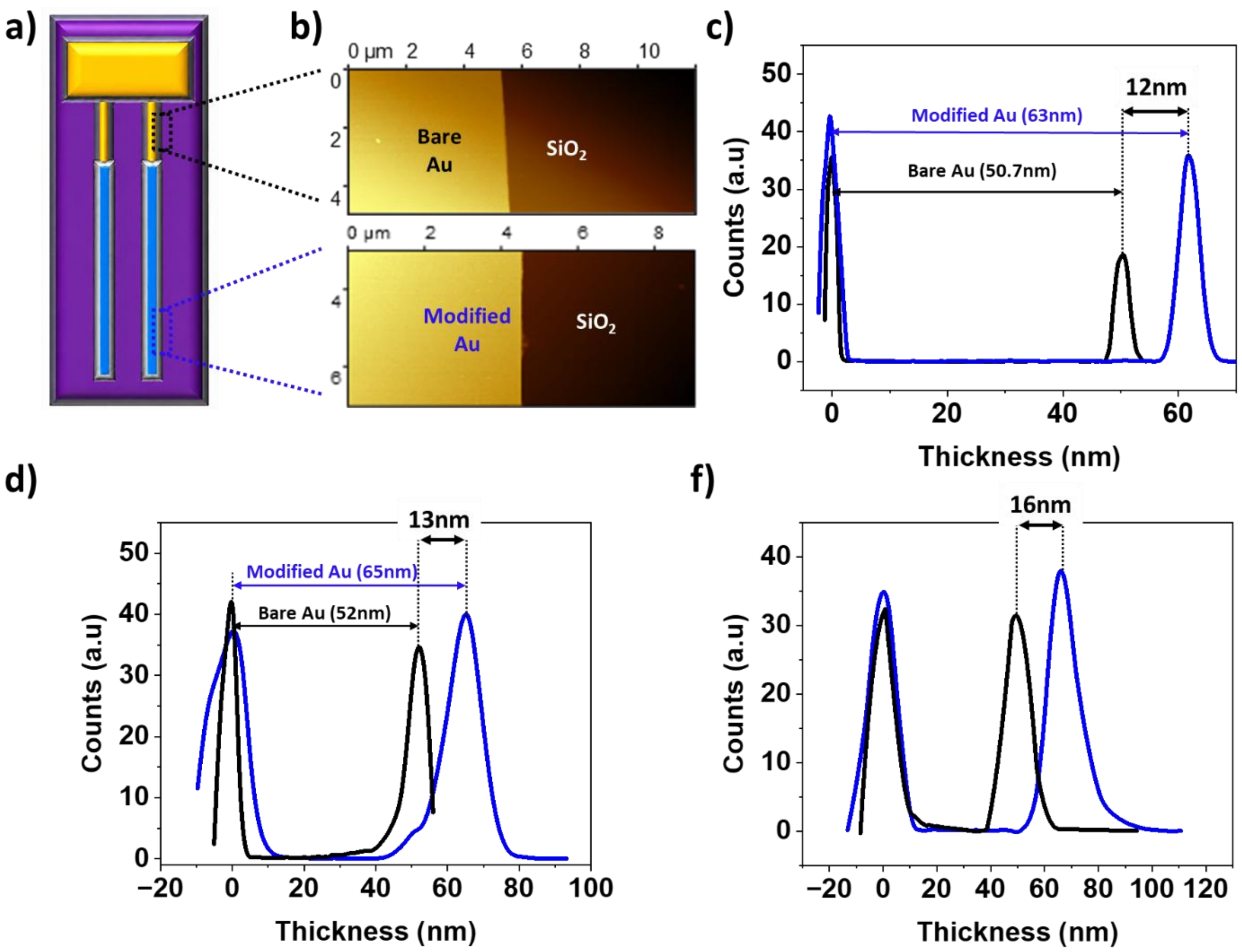

**Figure S4.** (a) Schematic representation of a microelectrode substrate consisting of two Au strips grafted from the bottom. (b) AFM images of the bare Au electrode and Si substrate separated by a step (top), and of the DAEVIO-functionalized Au electrode and Si substrate separated by a step (bottom). Gaussian statistical analysis of the corresponding AFM height distributions, showing DAEVIO film thicknesses of (c) 12 nm, (d) 13 nm, and (f) 16 nm.

| Samples | Substrate | Electrochemical conditions | Thickness (nm) | Measurement |
|---|---|---|---|---|
| **Sample 1** | gold stripe | [+0.2 -0.4]; 15 cycles; 100mV/s | 12±1 | Electrical measurements at ambient temperature |
| **Sample 2** | gold stripe | [+0.2 -0.4]; 15 cycles; 100mV/s | 12±1 | Electrical measurements at ambient temperature |
| **Sample 3** | gold stripe | [+0.2 -0.4]; 15 cycles; 100mV/s | 13±1 | Electrical measurements at ambient temperature |
| **Sample 4** | gold stripe | [+0.4 -0.4]; 15 cycles; 100mV/s | 16±1 | Temperature-dependent electrical measurements |
| **Sample 5** | continuous gold | [+0.4 -0.4]; 15 cycles; 100mV/s | 7±1 | XPS measurements |

**Table S2.** Electrografting parameters of DAEVIO on gold: potential scan window, number of cycles, scan rate, and film thickness determined by AFM.

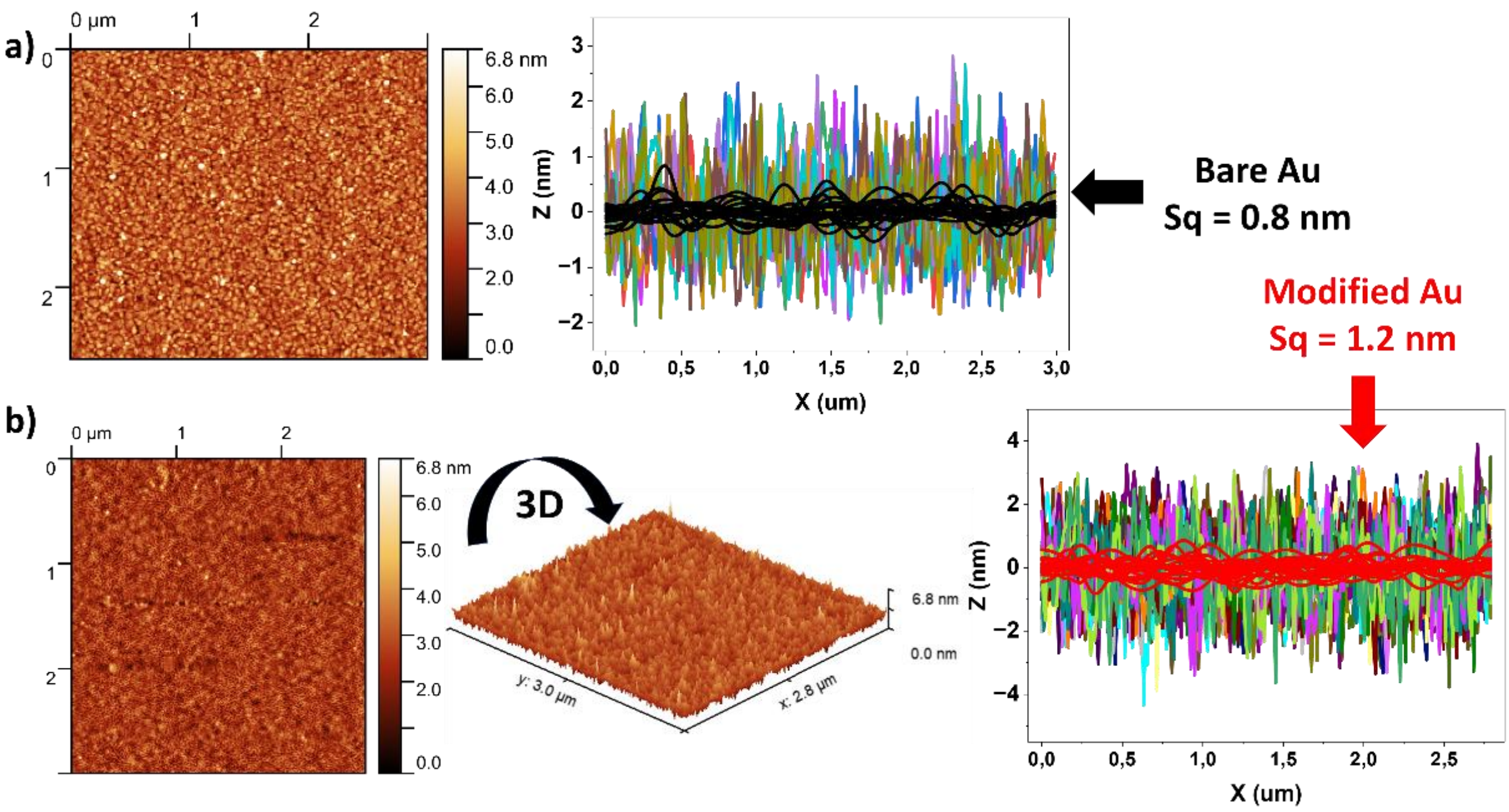

**Figure S5.** AFM topographic images and corresponding root-mean-square surface roughness (Sq) of the Au (50 nm) bottom electrode: (a) bare gold (Sq = 0.8 nm) and (b) DAEVIO-modified gold (Sq = 1.2 nm) (molecular thickness = 7nm).

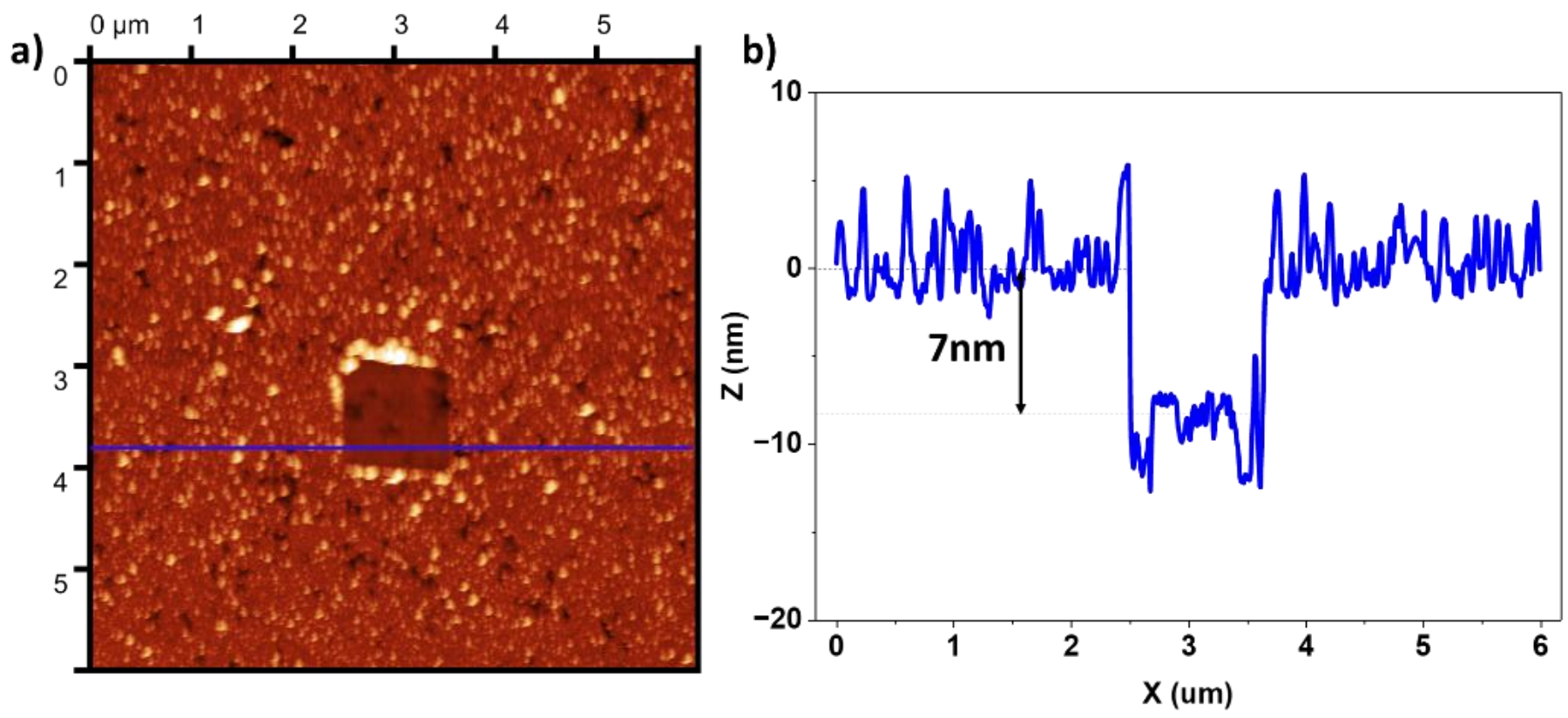

**Figure S6.** (a) AFM topography image performed in tapping mode of the surface modified with DAEVIO molecules used in the XPS measurements and subsequently scratched with the AFM tip (square region at the center of the image). (b) The corresponding line profile showing a molecular film thickness of approximately 7 nm.

## IV.2. XPS

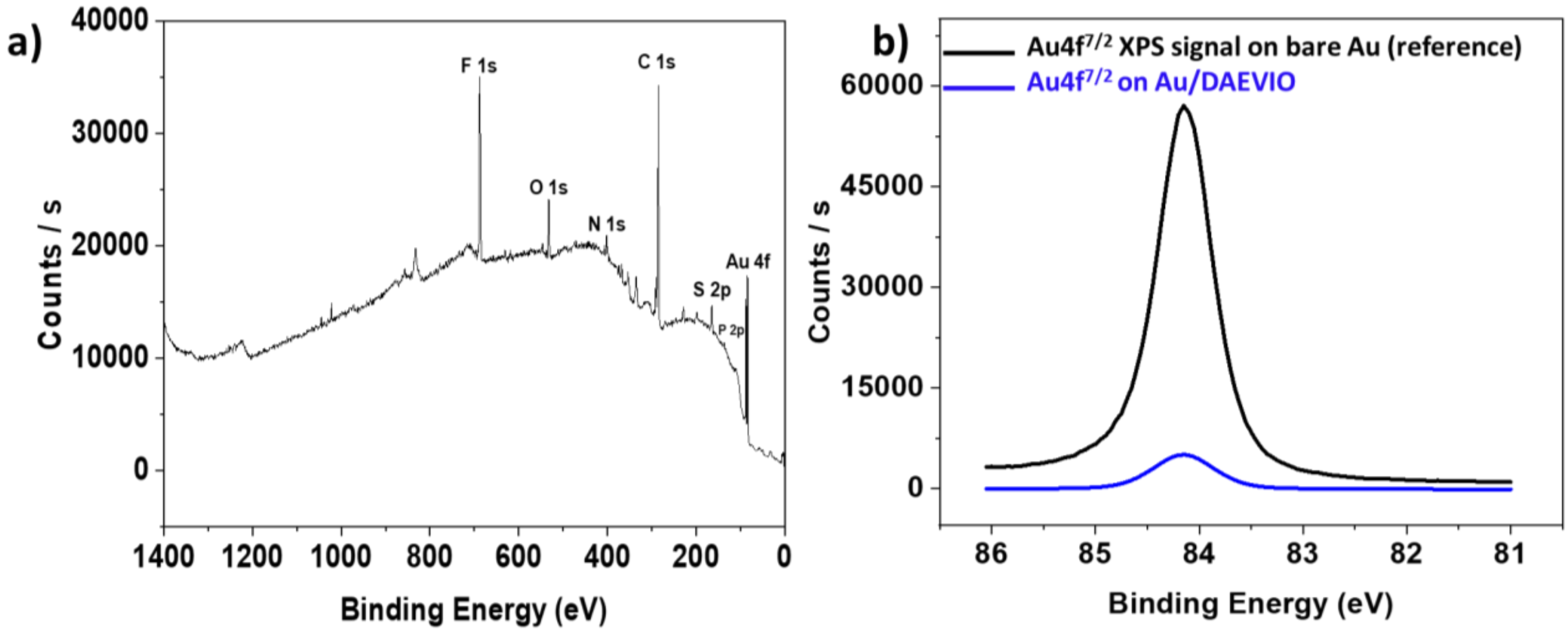

**Figure S7.** (a)The survey XPS spectrum of the DAEVIO-functionalized gold surface covering the full binding energy range. (b) High-resolution Au 4f XPS spectra of the bare gold surface (black) and the DAEVIO-functionalized gold surface (bleu).

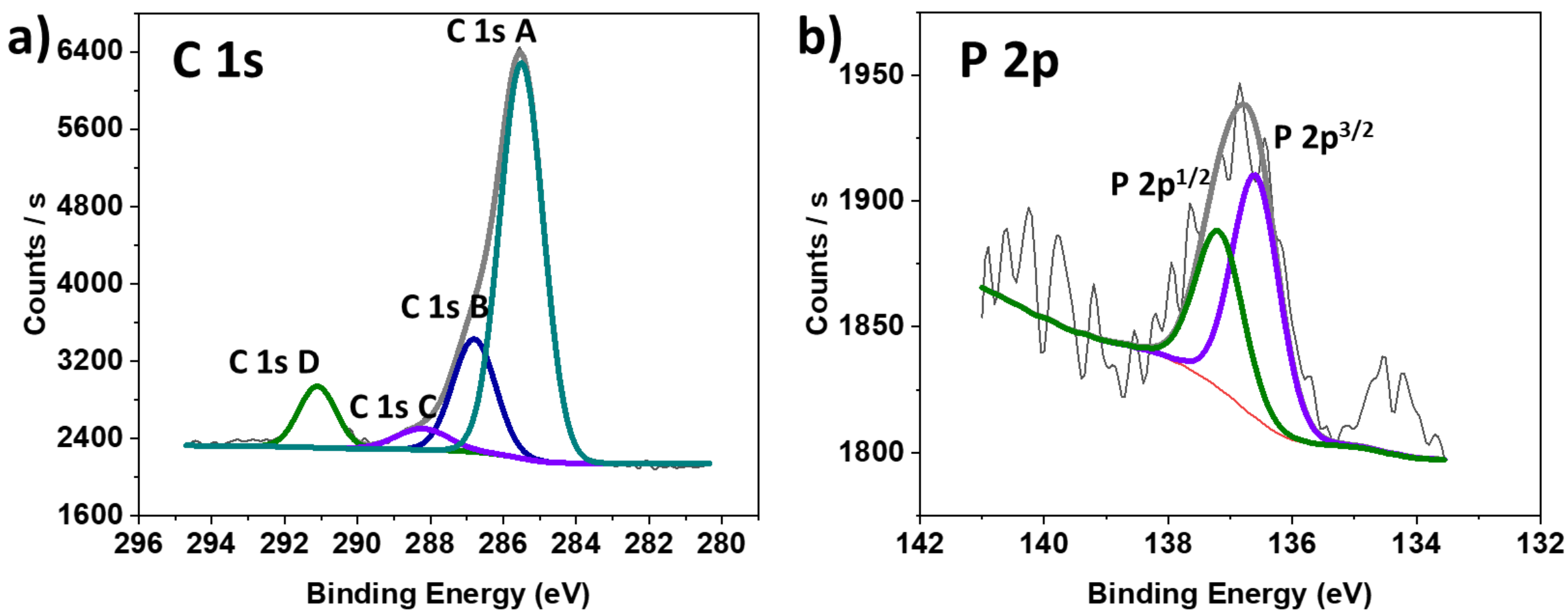

**Figure S8.** High-resolution XPS spectra of the corresponding DAEVIO-functionalized gold surface (a) P 2p and (b) C 1s.

| | Theoretical ratio ($VIO^{2+}$) | Theoretical ratio ($VIO^{•+}$) | Theoretical ratio ($VIO^{0}$) | Experimental ratio |
|---|---|---|---|---|
| **C/F** | 2.1 | **3.2** | 6.3 | **3.6** |
| **P/S** | 1 | **0.5** | 0 | **0.4** |

**Table S3.** Comparison between the theoretical and experimental atomic ratios determined from the corrected XPS peak areas.

## V. Light irradiation

To quantify the irradiation conditions, the optical power transmitted through the top electrode was measured using the 385 LED and an optical power meter on both a bare quartz substrate and a quartz/Ti (2nm)/Au (10nm) sample representative of the top electrode used in the molecular junctions. From the measured optical power and the illuminated spot area (3 mm diameter, corresponding to approximately 0.071 $cm^2$), the irradiance was calculated, together with the effective optical power received by a single 20 x20 $\mu m^2$ molecular junction. The corresponding values are summarized in **Table S4**.

At 385 nm, the optical power transmitted through the quartz/Ti (2nm)/Au (10nm) sample corresponds to approximately 34.3% of that measured through bare quartz. The incident irradiance decreases from 198.1 $mW \cdot cm^{-2}$ on bare quartz to 67.9 $mW \cdot cm^{-2}$ after transmission through the Ti (2nm)/Au(10nm) electrode. The corresponding optical power incident on a 20 x 20 μm$^2$ junction is approximately 0.272 μW.

| **λ (nm)** | **Substrate** | **Measured power (mW)** | **Irradiance (mW·cm$^{-2}$)** | **Power on a 400 μm$^2$ junction (μW)** | **Transmission through top electrode (%)** |
|---|---|---|---|---|---|
| **385** | bare quartz | 14.0 | 198.10 | 0.792 | 100.0 |
| **385** | quartz/$Ti_{2nm}Au_{10nm}$ | 4.8 | 67.90 | 0.272 | 34.3 |

**Table S4.** Optical power measured under 385 nm illumination using a bare quartz substrate and a quartz/Ti (2 nm)/Au (10 nm) sample, together with the corresponding irradiance and effective optical power incident on a single molecular junction with an active area of 20 x 20 μm$^2$ (illumination spot diameter = 3 mm).

To confirm the optical transmission of the top electrode, the transmission spectra of a bare quartz substrate (blue curve), an quartz/Ti (2nm)/Au (15nm) sample (dashed orange curve), and an quartz/Ti (2nm)/Au (10nm) sample (solid orange curve) were compared by UV-visible spectroscopy, as shown in **Figure S9**. The bare quartz substrate exhibits a transmission close to 100%. After deposition of the Ti/Au layers, the transmission decreases over the entire spectral range due to the reflection introduced by the metallic layers. For the quartz/Ti (2nm)/Au (15nm) sample, the transmission is approximately 22%, whereas it increases to about 34% for the quartz/Ti (2nm)/Au (10nm) sample (at 385nm). These results are in good agreement with the literature [4]. Consequently, the quartz/Ti (2nm)/Au (10nm) top electrode provides a suitable compromise between electrical conductivity and optical transparency, allowing sufficient light to reach the molecular layer while maintaining efficient electrical contact.

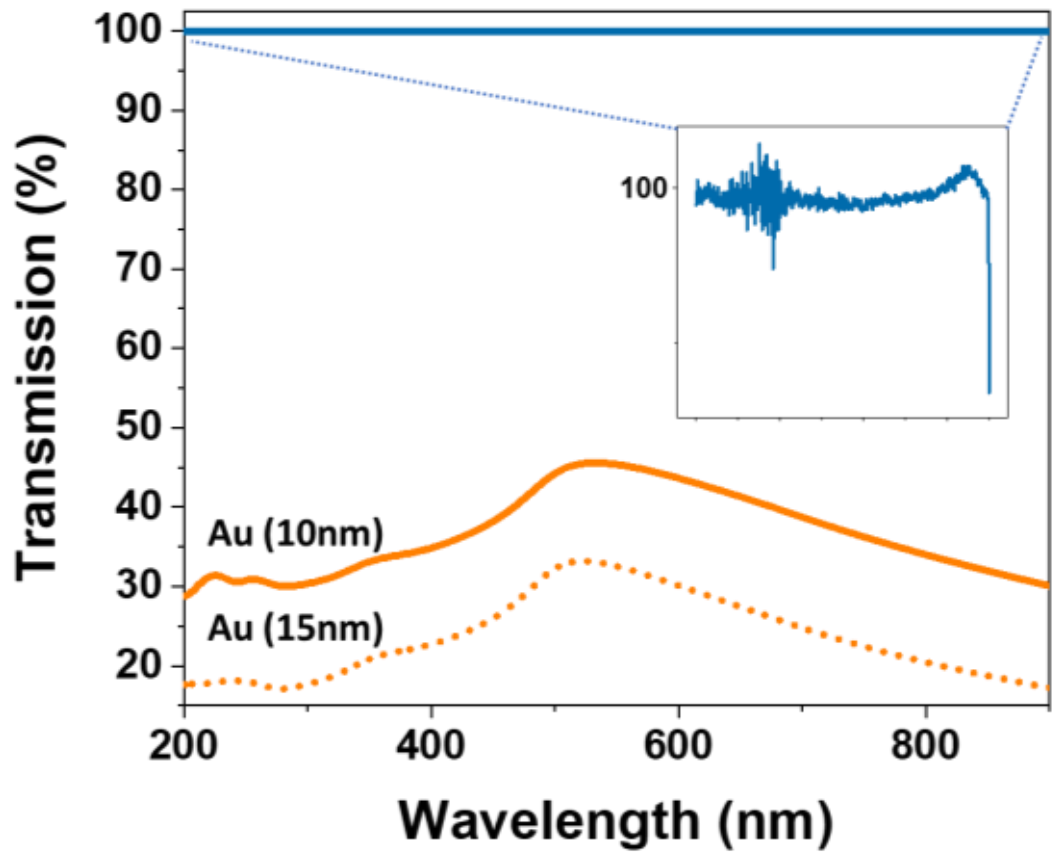

**Figure S9.** Optical transmission spectra of bare quartz used as a reference (blue curve) and quartz modified with Ti (2nm)/Au (15nm) or Ti (2nm)/Au (10nm) (orange curves).

# VI. Electrical characterization

## VI.1. Temperature-dependent electrical measurements of DAEVIO

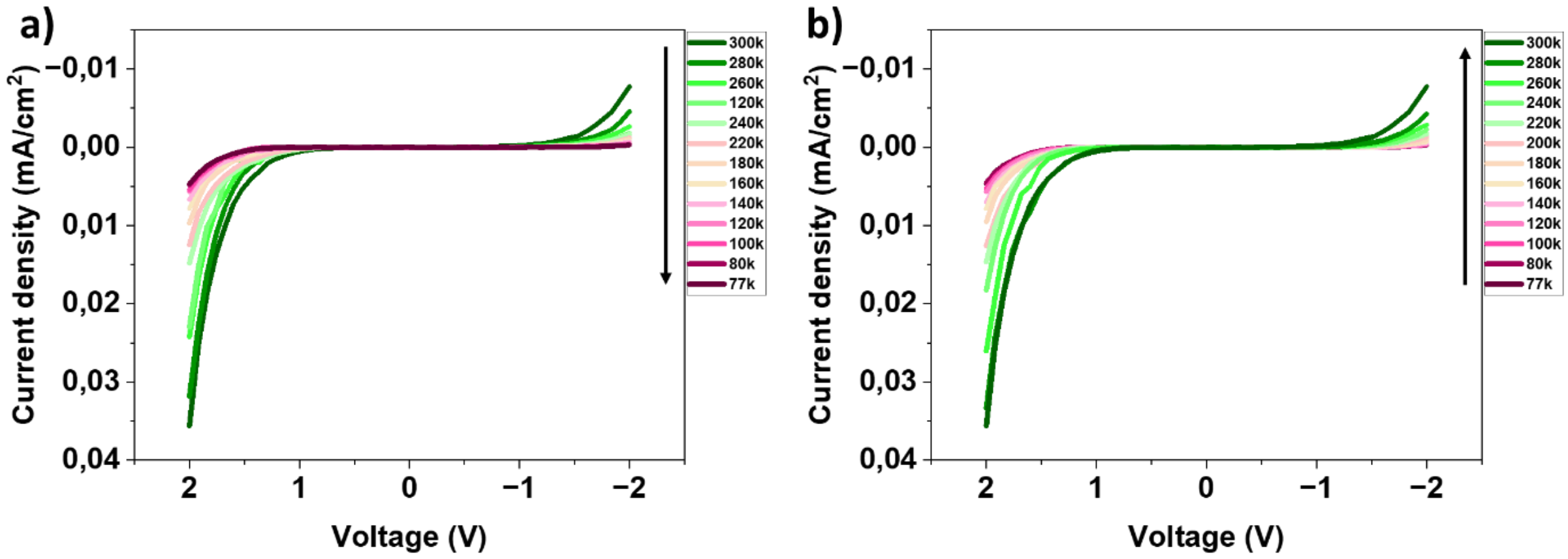

**Figure S11.** Temperature-dependent JV characteristics of Au (50nm)/DAEVIO (16nm)/Ti (2nm) Au (50nm) molecular junction in the OFF-state during (a) cooling from 300 to 77 K and (b) warming from 77 to 300 K, respectively.

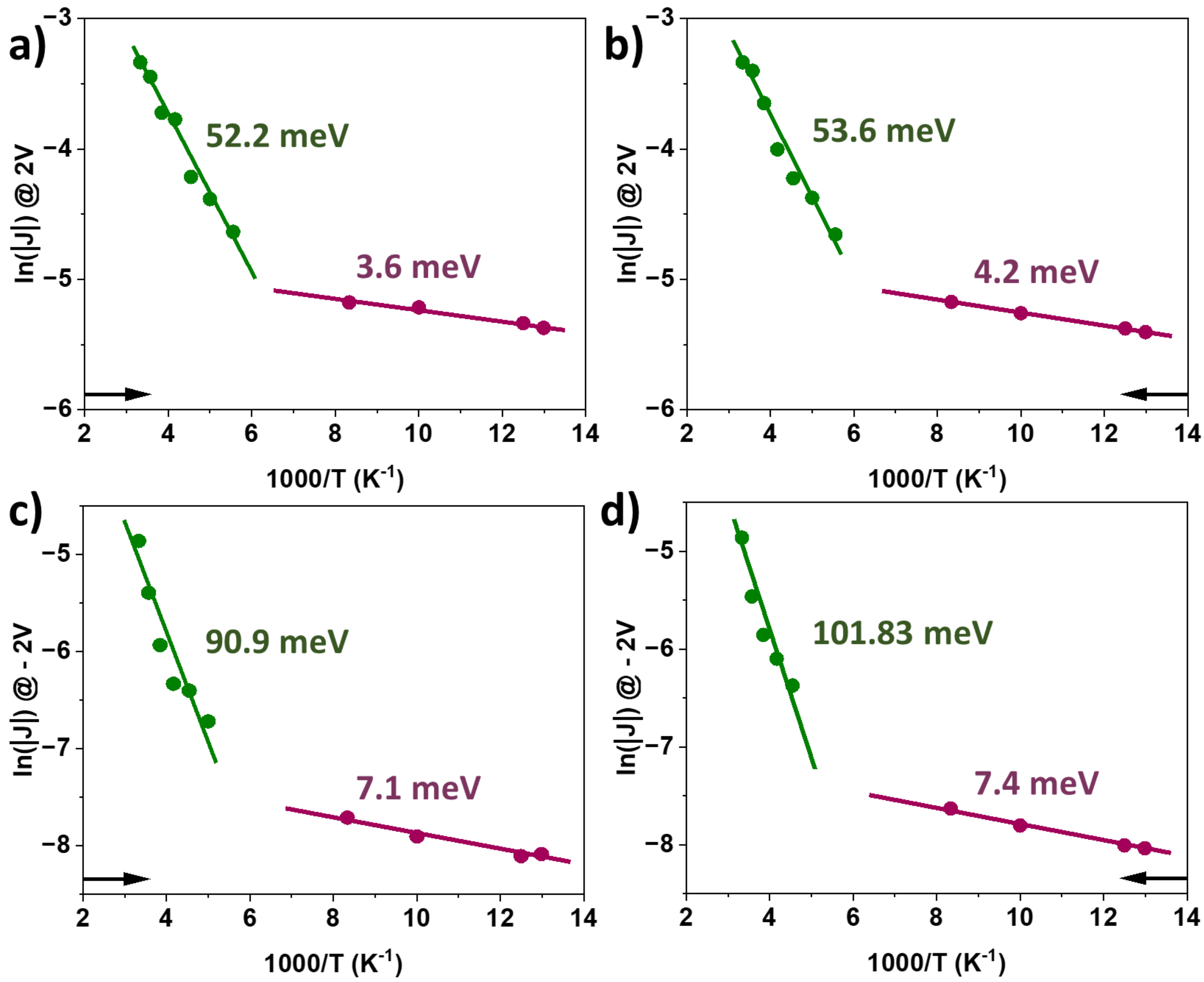

**Figure S12.** Arrhenius analysis (ln |J| vs. 1000/T) for OFF-state junction presented in **Figure S11** (a) +2V during cooling, $R^2$ = 0.97, (b) +2 V during warming, $R^2$ = 0.96 and 0.99, (c) -2V during cooling, $R^2$ = 0.96 and 0.98 and (d) -2V during warming, $R^2$ = 0.92 and 0.99. Green and magenta lines represent linear fits over the high-temperature (300-180 K) and low-temperature (160-77 K) intervals, respectively. The similarity between the cooling and warming data indicates a reversible electrical response within the experimental resolution.

## VI.2. EQE and IQE

The magnitude of the zero-bias photocurrent was used to estimate the external quantum efficiency. The EQE was calculated from EQE = $\frac{Jsc*h*c}{q*Pi*\lambda}$ , where Pi is the incident optical power density, h Planck's constant, c the speed of light, q the elementary charge and λ = 385 nm. At this wavelength, when J and Pi are expressed in the same mA/mW units, EQE= 3.22 $\frac{Jsc}{Pi}$. Using JSC = 0.006 mA $cm^{-2}$ and Pi = 198.1 mW $cm^{-2}$, we obtain EQE=$10^{-4}$, or 0.01%. This value represents the number of collected electrons relative to the number of photons incident on the device. It is relatively low compared with optimized photoactive molecular bilayer junctions,[8,9] indicating that only a small fraction of incident photons ultimately contributes to the collected short-circuit current.

Part of this loss originates from the limited optical transmission of the top electrode. The Ti (2 nm Au10 nm) electrode transmits T = 34.3% of the incident light at 385 nm (**see Table S4** and **Figure S9**). Correcting the incident photon flux only for this transmission gives IQE' = $\frac{EQE}{T}$ and therefore IQE' = $3*10^{-4}$, or 0.03%. We use IQE' here to distinguish this transmission-corrected value from a true internal quantum efficiency IQE. The fraction of transmitted photons actually absorbed by the 12 nm DAEVIO layer was not measured and is therefore not included in the calculation.